**Thermodynamics of mixtures with strong negative deviations from Raoult's law. XVIII: Excess molar enthalpies for the (1-alkanol + cyclohexylamine) systems at 298.15 K and modelling**

Luis Felipe Sanz,[a] Juan Antonio González,[a*] Isaías. García de la Fuente,[a] José Carlos Cobos[a] and Fernando Hevia[b]

[a]G.E.T.E.F., Departamento de Física Aplicada, Facultad de Ciencias, Universidad de Valladolid, Paseo de Belén, 7, 47011 Valladolid, Spain.

[b]Université Clermont Auvergne, CNRS. Institut de Chimie de Clermont-Ferrand. F-63000, Clermont-Ferrand, France b Departamento de Física Aplacada. EIFAB. Campus D

*corresponding author, e-mail: jagl@termo.uva.es; Fax: +34-983-423136; Tel: +34-983-423757

**Abstract**

Excess molar enthalpies, $H_m^E$, have been measured using a Tian-Calvet microcalorimeter for the systems $CH_3(CH_2)_{u-1}OH$ ($u$ =1,2,3,4,7,10) + cyclohexylamine at 298.15 K and 0.1 MPa. The $H_m^E$ values are large and negative, indicating the existence of strong interactions between unlike molecules, which is consistent with the also large and negative excess molar volumes, $V_m^E$ of these solutions, previously measured by us. The contribution from the equation of state term to $H_m^E$ has been evaluated for the 1-alkanol + cyclohexylamine, or + 1-hexylamine, or + aniline mixtures, and the corresponding excess molar internal energies at constant volumes, $U_{m,V}^E$, determined. It is shown that such contribution is particularly important for the methanol + aniline system, in such way that the excess functions $H_m^E$ and $U_{m,V}^E$ have different sign at $x_1$ = 0.5. The DISQUAC and ERAS models have been applied to the cyclohexylamine systems, and the interaction parameters reported. DISQUAC improves ERAS results on $H_m^E$. The latter model describes correctly the $V_m^E$ curves. The variation of $H_m^E$ of $CH_3(CH_2)_{u-1}OH$ + cyclohexylamine, or + 1-hexylamine, or + aniline mixtures with $u$ along a homologous series with a given amine, or with the amine in mixtures with a given 1-alkanol is discussed in terms of the different interactional contributions to $H_m^E$.

Keywords: 1-alkanol; cyclohexylamine; $H_m^E$; $V_m^E$; DISQUAC; ERAS

1. **Introduction**

Alkanols and amines have many industrial applications. Alkanols are used in the manufacturing of fuels, perfumes, cosmetics, paints, drugs, explosives, fats, waxes, resins, plastics and more [1,2]. Cyclic amines are important in the production of pharmaceuticals and of chemicals which are included in insecticides, pesticides, dyes, or corrosion inhibitors [3]. The study of 1-alkanol + amine mixtures is required, *e.g.*, to a better understanding of solutions containing alkanolamines that are used in the chemical absorption of $CO_2$ emissions [4].

Over the last years, we have conducted a systematic research on mixtures containing amines, and, particularly on 1-alkanol + amine systems. This class of solutions is very interesting since they show rather different behaviours. For example, mixtures with linear or primary amines are characterized by strong negative deviations from the Raoult's law as it is demonstrated by their large and negative excess molar Gibbs energies ($G_m^E$) [5-10] and enthalpies, ($H_m^E$), [11-18]. Thus, for the methanol + 1-butylamine mixture at 348.15 K and equimolar composition, $G_m^E = -799$ J·mol$^{-1}$ [6] and for methanol + 1-hexylamine system at 298.15 K and $x_1 = 0.5$, $H_m^E = -3200$ J·mol$^{-1}$ [11]. These results reveal the existence of strong interactions between unlike molecules [11,12,18], which are even stronger than those between 1-alkanol molecules, and that lead to the formation of complexes as solid-liquid phase diagrams reveal [19,20]. In contrast, 1-alcohol + aniline mixtures show positive deviations from the Raoult's law as it can be seen from the following results for the ethanol system at 298.15 K and equimolar composition: $G_m^E = 515$ J·mol$^{-1}$ [21] and $H_m^E = 360$ J·mol$^{-1}$ [22]. Up to now, we have provided volumetric [23-30], calorimetric [31-33], phase equilibria [10,34], viscosimetric [27-29] and permittivity [27-30,35-37] data for binary mixtures formed by 1-alkanol and any of the following amines: linear primary or secondary amines, *N,N,N*-triethylamine, cyclohexylamine, aniline, benzylamine, quinoline or 1-*H*-pyrrole. We have also developed detailed theoretical investigations on this type of systems [35-44] by means of different models: DISQUAC [45], ERAS [46], Flory [47], Kirkwood-Buff integrals formalism [48,49], the concentration-concentration structure factor formalism [50,51], or the Kirkwood-Fröhlich theory [52]. As continuation of our works on 1-alkanol + cyclohexylamine systems [27-29], we report now $H_m^E$ data, at 298.15 K and 0.1 MPa for systems involving methanol, ethanol, 1-propanol, 1-butanol, 1-heptanol or 1-decanol. In addition, these mixtures are studied in terms of the DISQUAC and ERAS models. Previously, we had provided ERAS parameters determined essentially from excess molar volume, $V_m^E$, data [29]. No interaction parameters for the present solutions are available in the framework of the UNIFAC (Dortmund) model [53,54].

## 2. Experimental

### 2.1 Materials

All the chemicals were supplied by Sigma-Aldrich. Their CAS number, purity, according to gas chromatographic analysis (GC) provided by the supplier, and densities are shown in Table 1. Densities of the pure liquids were determined from a vibrating-tube densimeter Anton Paar model DSA 602 with a temperature stability of 0.01 K. Calibration of the densimeter was conducted using the following pure liquids: heptane, 2,2,4-trimethylpentane, cyclohexane, benzene, toluene, 1-propanol and water. The estimated uncertainty for density is $\pm$ 8x10$^{-2}$ kg·m$^{-3}$.

### 2.2 Apparatus and procedure

Compounds were weighed using an analytical balance A and D instrument model HR-202 (weighing uncertainty 0.1 mg), taking into account the corresponding corrections on buoyancy effects. The standard uncertainty in the final mole fraction is 0.0005. Molar quantities were calculated using the relative atomic mass Table of 2015 issued by the Commission on Isotopic Abundances and Atomic Weights (IUPAC) [55]. $H_m^E$ measurements were carried out at 298.15 K and 0.1 MPa by means of a standard Tian-Calvet microcalorimeter equipped with an aluminium mixing cell, designed by us, with a small (<2%) gas phase. The mixing process is the same as in previous applications [56]. Some improvements have been performed with regard to the thermal insulation and to the acquisition and processing of the data obtained from the apparatus. A new cable, a shielded twisted pair, has been used especially adapted for the type of thermocouples of the calorimeter. This allows reducing the appearance of parasitic voltages and noise in the output signal produced by external sources. The calibration of the calorimeter was conducted measuring $H_m^E$ for the cyclohexane + benzene, and methanol + 1-butylamine systems at 298.15 K and 0.1 MPa. Results are collected in Table 2. A comparison with data from the literature [16,18,57-62] is shown in Figures 1a and 1b. At equimolar composition, our measurements deviate by ~1% with regard to the corresponding values from [16,18,57-60]. The values provided by Pradhan and Mathur (−3866 J·mol$^{-1}$ [61]) and by Dutta-Choudhury and Mathur (−3850 J·mol$^{-1}$ [62]) for the methanol + 1-butylamine mixture are lower than our result (−3726 J·mol$^{-1}$) by 3.7% and 3.3%, respectively. The estimated maximum relative uncertainty for $H_m^E$ is 0.015. Since the systems selected for the calibration show very different $H_m^E$ values (Table 2), this guarantees that the equipment is useful to perform $H_m^E$ measurements over a wide range of values.

### 3. Experimental results

Our measurements on $H_m^E$ for 1-alkanol + cyclohexylamine mixtures at 298.15 K and 0.1 MPa are listed in Table 3 (see Figures S1 and S2, supplementary material). Measurements of $H_m^E$ for the methanol system have been reported by Mato and Berrueta [60]. At $x_1 = 0.5$, their value is ca. 600 J·mol$^{-1}$ higher than our experimental result (Figure 2). Partial excess molar enthalpies at infinite dilution of component i (=1,2), $H_{mi}^{E,\infty}$, are available in the literature for the methanol(1) + cyclohexylamine(2) mixture [63]. The results are: $H_{mi}^{E,\infty}$/kJ.mol$^{-1}$ = –11.55 (i = 1); –14.90 (i = 2) [63]. Our values ($H_{mi}^{E,\infty}$/kJ·mol$^{-1}$ = –11.1 (i = 1); –14.1 (i = 2)), determined from $H_m^E$ measurements over the entire mole fraction range are in good agreement with the mentioned results. Excess molar internal energies of constant volume, $U_{m,V}^E$, can be determined from [64]:

$$U_{m,V}^E = H_m^E - T\frac{\alpha_p}{\kappa_T}V_m^E \tag{1}$$

where $\alpha_p$ and $\kappa_T$ are, respectively, the isobaric thermal expansion coefficient and the coefficient of isothermal compressibility of the considered system. The $U_{m,V}^E$ values for 1-alkanol + cyclohexylamine mixtures were obtained (see Table S1, supplementary material) using the $H_m^E$ values listed in Table 3 and $V_m^E$ results at the same mole fractions determined from Redlich-Kister expansions previously obtained when volumetric properties for these solutions were measured [27-29]. The $\alpha_p$ and $\kappa_T$ values were calculated assuming ideal behavior for the systems ($M^{id} = \phi_1 M_1 + \phi_2 M_2$; with $M_i = \alpha_{pi}$, or $\kappa_{T_i}$ and $\phi_i = x_i V_{m,i}/(x_1 V_{m,1} + x_2 V_{m,2})$). For pure compounds, their $\alpha_{pi}$, and $\kappa_{T_i}$ values were taken from the literature [65-67]. The contribution of the equation of state (eos) term to $H_m^E$, defined as $T\frac{\alpha_p}{\kappa_T}V_m^E$, is ranged between 17% for the methanol mixture and 12% for the 1-decanol solution. However, the mentioned contribution may be much more important, as it will be seen below.

The data ($F_m^E = H_m^E$; $F_m^E = U_{m,V}^E$) were fitted by unweighted least-squares polynomial regression to the equation of the Redlich-Kister type [68]:

$$F_m^E = x_1(1-x_1)\sum_{i=0}^{k-1} A_i(2x_1-1)^i \tag{2}$$

The number, $k$, of needed coefficients for this regression was determined, for each system, by applying an F-test of additional term [69] at 99.5% confidence level. For cyclohexylamine mixtures, Table 4 lists the parameters $A_i$ obtained in the regression, together with the standard deviations $\sigma(F_m^E)$ defined by:

$$\sigma(F_m^E) = \left[\frac{1}{N-k}\sum_{j=1}^{N}\left(F_{mcal,j}^E - F_{mexp,j}^E\right)^2\right]^{1/2} \tag{3}$$

where $N$ stands for the number of data points, and $F_{mcal,j}^E$ is the value of the excess property calculated using equation (2). For the systems cyclohexane + benzene, and methanol + 1-butylamine, this information is given as a footnote in Table 2.

## 4. Models

### *4.1 DISpersive QUAsiChemical*

DISQUAC is based on the rigid lattice theory developed by Guggenheim [70]. Some important features of the model are given. (i) The geometrical parameters: total molecular volumes, $r_i$, surfaces, $q_i$, and the molecular surface fractions, $\alpha_{si}$, of the mixture components are calculated additively using the group volumes $R_G$ and surfaces $Q_G$ recommended by Bondi [71], with the volume $R_{CH4}$ and surface $Q_{CH4}$ of methane taken arbitrarily as volume and surface units [72]. For the groups involved in this investigation, the geometrical parameters are available in the literature [72-74] (ii) The partition function is factorized into two terms. The excess functions $G_m^E$ and $H_m^E$ are the result of the sum of two contributions. The dispersive (DIS) term is linked to the contribution from dispersive forces; and the quasichemical (QUAC) term is due to the anisotropy of the field forces created by the solution molecules. In the case of $G_m^E$, a combinatorial term, $G_m^{E,COMB}$, calculated using the Flory-Huggins equation [72,75] must be included. Therefore,

$$G_m^E = G_m^{E,DIS} + G_m^{E,QUAC} + G_m^{E,COMB} \tag{4}$$

$$H_m^E = H_m^{E,DIS} + H_m^{E,QUAC} \tag{5}$$

(iii) The interaction parameters are dependent on the molecular structure of the mixture components; (iv) In the present status of the theory, it is not possible to characterize each polar contact by its own coordination number ($z$). For this reason, $z = 4$ is used for all the polar contacts. This is a shortcoming of DISQUAC and is partially removed assuming that the interaction parameters are dependent on the molecular structure. (v) It is also assumed that $V_m^E = 0$.

The equations used to calculate the DIS and QUAC contributions to $G_m^E$ and $H_m^E$ can be found elsewhere [38,76]. The temperature dependence of the interaction parameters is expressed in terms of the DIS and QUAC interchange coefficients [38,76], $C_{st,l}^{DIS}; C_{st,l}^{QUAC}$ where s ≠ t are two contact surfaces present in the mixture and $l$ = 1 (Gibbs energy; $C_{st,1}^{DIS/QUAC} = g_{st}^{DIS/QUAC}(T_o)/RT_o$); $l$ = 2 (enthalpy, $C_{st,2}^{DIS/QUAC} = h_{st}^{DIS/QUAC}(T_o)/RT_o$)), $l$ = 3 (heat capacity, $C_{st,3}^{DIS/QUAC} = c_{pst}^{DIS/QUAC}(T_o)/R$)). $T_o$ = 298.15 K is the scaling temperature and $R$, the gas constant.

*4.2 ERAS*

The Extended Real Associated Solution (ERAS) model [46] combines the Real Association Solution Model [77,78] with Flory's equation of state [47]. We provide some relevant features of the model. (i) The excess molar functions ($F_m^E = H_m^E$, $V_m^E$) are calculated as the sum of two contributions. The chemical contribution, $F_{m,chem}^E$, arises from hydrogen bonding; the physical contribution, $F_{m,phys}^E$, is linked to nonpolar Van der Waals interactions and free volume effects. Expressions for $H_m^E$ and $V_m^E$ can be found elsewhere [38]. (ii) It is assumed that only consecutive linear association occurs. Accordingly, self-association is described by a chemical equilibrium constant ($K_i$) independent of the chain length of the self-associated species A or B (in this case, i = A (1-alkanol) or = B (cyclohexylamine)), according to the equations:

$$A_m + A \xrightleftharpoons{K_A} A_{m+1} \quad (6)$$

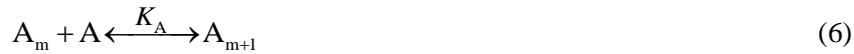

$$B_n + B \xrightleftharpoons{K_B} B_{n+1} \quad (7)$$

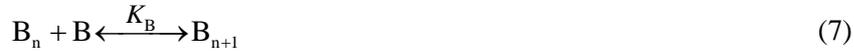

with m and n ranging from 1 to $\infty$. The cross-association between two self-associated species $A_m$ and $B_n$ is represented by:

$$A_m + B_n \xrightleftharpoons{K_{AB}} A_m B_n \quad (8)$$

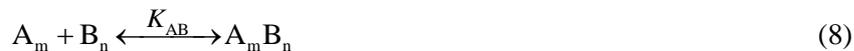

where cross-association constants, $K_{AB}$, are also considered to be independent of the chain length. Reactions described by equations (6)-(8) are characterized, respectively, by the molar enthalpies of intermolecular hydrogen-bonding $\Delta h_A^*$, $\Delta h_B^*$ and $\Delta h_{AB}^*$, and by negative molar hydrogen-bonding volumes, $\Delta v_A^*$, $\Delta v_B^*$ and $\Delta v_{AB}^*$, defined in order to take into account the decrease of the core volume of multimers in comparison to that of an isolated monomer. The three equilibrium constants depend on temperature according to the $\Delta h_i^*$ values and the Van't Hoff equation. (iii) The $F_{m,phys}^E$ term is derived from the Flory's equation of state [47], which is assumed to be valid not only for pure compounds but also for the mixture [11,18]:

$$\frac{\bar{p}_i \bar{V}_i}{\bar{T}_i} = \frac{\bar{V}_i^{1/3}}{\bar{V}_i^{1/3} - 1} - \frac{1}{\bar{V}_i \bar{T}_i} \tag{9}$$

where $i$ = A, B or M (mixture). In equation (9), $\bar{V}_i = V_{m,i}/V_{m,i}^*$; $\bar{p}_i = p/p_i^*$; $\bar{T}_i = T/T_i^*$ are the reduced properties for volume, pressure and temperature, respectively. The pure component reduction parameters ($V_{m,i}^*$, $p_i^*$, $T_i^*$) are obtained from $p$-$V$-$T$ data (density, $\alpha_{pi}$, and $\kappa_{T_i}$) and association parameters [11,18]. The reduction parameters for the mixture $p_M^*$ and $T_M^*$ are calculated from mixing rules [11,18]. The total relative molecular volumes and surfaces of the compounds were calculated additively on the basis of the group volumes and surfaces recommended by Bondi [71].

## 5. Adjustment of interaction parameters

### 5.1 DISQUAC

In terms of DISQUAC, 1-alkanol + cyclohexyalmine mixtures are built by four types of surface: (i) type a, aliphatic (CH$_3$, CH$_2$, in 1-alkanols); (ii) type c, cyclic (c-CH$_2$ or c-CH in cyclohexylamine; (iii) type h, OH in 1-alkanols; (iv) type n, amine (NH$_2$ in cyclohexylamine). The four surfaces generate six contacts: (a,c), (a,h), (a,n), (c,h), (c,n) and (h,n). The (a,c) contact is represented by dispersive interaction parameters obtained from the study of cyclohexane + $n$-alkane mixtures [75]. The remainder contacts are described by both DIS and QUAC interaction parameters. The interchange coefficients $C_{an,l}^{DIS}$ and $C_{an,l}^{QUAC}$ are known from a general DISQUAC treatment of cyclic amine + alkane systems [79], and the $C_{sh,l}^{DIS}$ and $C_{sh,l}^{QUAC}$ coefficients from the corresponding investigation of 1-alkanol + $n$-alkane (s = a) [74], or + cyclohexane (s = c) [80] mixtures. Therefore, only the interaction parameters for the (h,n) contact must be determined.

The general procedure applied in the estimation of the interaction parameters has been explained in detail elsewhere [38,76]. Due to the VLE data for these systems is scarce [81,82], the $C_{hn,1}^{QUAC}$ coefficients (l =1,2) were adjusted together with the $C_{hn,2}^{DIS}$ coefficients to get a good description of the symmetry of the $H_m^E$ curves. The first DIS Gibbs energy parameters were then estimated using the few VLE available in the literature [81,82]. Final parameters are listed in Table 5.

5.2  *ERAS*

The values of the ERAS parameters for 1-alkanols, $K_A$, $\Delta h_A^*$, $\Delta v_A^*$ and for cyclohexylamine, $K_B$ $\Delta h_B^*$, $\Delta v_B^*$ are known from the study of the corresponding mixtures with alkanes [46,83]. The binary parameters to be fitted to the $H_m^E$ and $V_m^E$ data of the systems studied are then $K_{AB}$, $\Delta h_{AB}^*$, $\Delta v_{AB}^*$ and $X_{AB}$ (Table 6).

## 6  Theoretical results

Comparison between experimental $H_m^E$ data and theoretical results using DISQUAC and ERAS is shown Table 7 (see also Figures 2 and 3 and Figures S3 and S4 of supplementary material). We note that DISQUAC improves results from the ERAS model. Larger differences with the experimental values emerge for systems with 1-heptanol or 1-decanol (Table 7). ERAS correctly describes $V_m^E$ (see Figure 3 and Figure S4 of supplementary material). On the other hand, DISQUAC provides, at equimolar composition and 407.1 K, $G_m^E$/J·mol$^{-1}$ = $-476$, (methanol); $-460$ (ethanol and $-294$ (1-propanol). The experimental results, in the same units, are respectively: $-508$, $-408$ and $-503$ [82]. This experimental variation of $G_m^E$ with the alkanol size should be taken with caution.

## 7.  Discussion

Below, we are referring to values of the thermodynamic properties at equimolar composition and 298.15 K. On the other hand, $n_{OH}$ stands for the number of C atoms of the 1-alkanol.

Figure 4 shows $H_m^E$ and $U_{m,V}^E$ values for 1-alkanol + amine mixtures. The $U_{m,V}^E$ values for systems with 1-hexylamine or aniline were computed by means of the same method explained above, using values of $\alpha_{pi}$, and $\kappa_{T_i}$ from the literature [30, 65-67].

As already mentioned, the large and negative $H_m^E$ values of 1-alkanol + 1-hexylamine mixtures (Figure 4) are due to the existence of strong interactions between unlike molecules [11,18,38]. On the other hand, $H_m^E$ increases from $n_{OH}$ = 1 up to $n_{OH}$ = 3, and then increases smoothly (Figure 4). The corresponding $V_m^E$ values are also negative (Figure 5), and it reveals that this excess function is determined mainly by interactional effects. The replacement of 1-hexylamine by cyclohexylamine (cyclization effect) in systems with a given 1-alkanol leads to decreased values of $H_m^E$ and $V_m^E$ when $n_{OH} \leq 4$ (Figures 4 and 5). For mixtures including longer

1-alkanols, $H_m^E$ and $U_{m,V}^E$ are more or less independent of the amine, while $V_m^E$ becomes higher for systems with cyclohexylamine (Figures 4 and 5). The latter may be ascribed to the cyclic amine breaks a larger number of alkanol-alkanol interactions (see below), and, in some extent, to the existence of increased free volume effects in 1-hexylamine solutions, as the corresponding ($\alpha_p/10^{-3}$) values of pure compounds suggest (in K$^{-1}$): 8.52 (1-octanol) [65]; 8.18 (1-decanol) [65]; 10.5 (cyclohexylamine) [67]; 11.28 (1-hexylamine) [84]. It is to be noted that $H_m^E$ and $V_m^E$ change with $n_{OH}$ in systems with cyclohexylamine more rapidly than in 1-hexylamine solutions (Figures 4 and 5). If 1-hexylamine is replaced by aniline (aromacity effect), the values of $H_m^E$ and $V_m^E$ increase in line with $n_{OH}$. For the methanol mixture, $H_m^E$ is negative (–175 J·mol$^{-1}$ [85]) while our $U_{m,V}^E$ result is positive (143 J·mol$^{-1}$). This huge variation underlines the importance of the eos contribution to $H_m^E$. In view of the positive $U_{m,V}^E$ values, one can conclude that interactions between like molecules are dominant in these mixtures. It is remarkable the opposite sign of the $U_{m,V}^E$ and $V_m^E$ values (Figures 4 and 5), that indicates that the contribution to $V_m^E$ arising from structural effects is here prevalent by far.

*7.1 The enthalpy of the 1-alkanol-amine interactions*

Next, we evaluate the enthalpy of the H-bonds between 1-alkanols and amines (termed as $\Delta H_{OH-NH2}^{int}$). If structural effects are neglected [64,86], $H_m^E$ can be considered the result of three contributions. Two of them, $\Delta H_{OH-OH}$, $\Delta H_{NH2-NH2}$, are positive, and arise, respectively, from the breaking of alkanol-alkanol and amine-amine interactions upon mixing. In this process, new OH---NH2 interactions are created, and it implies a negative contribution, $\Delta H_{OH-NH2}$, to $H_m^E$. Therefore [87-89]:

$$H_m^E = \Delta H_{OH-OH} + \Delta H_{NH2-NH2} + \Delta H_{OH-NH2} \qquad (10)$$

Values of $\Delta H_{OH-NH2}^{int}$ can be obtained extending the equation (10) to $x_1 \to 0$ [89-91]. Then, $\Delta H_{OH-OH}$ and $\Delta H_{NH2-NH2}$ can be replaced by $H_{m1}^{E,\infty}$ of 1-alkanol(1) or amine(1) + alkane(2) systems. Thus,

$$\Delta H_{OH-NH2}^{int} = H_{m1}^{E,\infty}(1-\text{alkanol} + \text{amine})$$
$$-H_{m1}^{E,\infty}(1-\text{alkanol} + \text{heptane}) - H_{m1}^{E,\infty}(\text{amine} + \text{heptane, or} + C_6H_{12}) \qquad (11)$$

Some shortcomings of this estimation of $\Delta H_{\text{OH-NH2}}^{\text{int}}$ values are now given. (i) $H_{\text{m1}}^{\text{E},\infty}$ data used were determined from $H_{\text{m}}^{\text{E}}$ measurements over the entire mole fraction range. (ii) For 1-alkanol + $n$-alkane systems, $H_{\text{m1}}^{\text{E},\infty}$ is assumed to be independent of the alcohol, a typical approach in the framework of association theories [46,92-94]. As in previous works [91,95], we have used here $H_{\text{m1}}^{\text{E},\infty}$ = 23.2 kJ·mol$^{-1}$ [96-98]. Nevertheless, it should be remarked that the values of $\Delta H_{\text{OH-NH2}}^{\text{int}}$ collected in Table 8 are still meaningful since they were obtained following the same procedure that in previous applications, which allows comparing enthalpies of interaction between 1-alkanols and different organic solvents. Inspection of Table 8 reveals that, for systems including 1-hexylamine or cyclohexylamine, $\Delta H_{\text{OH-NH2}}^{\text{int}}$ values are quite similar, although, in the latter mixtures, interactions between unlike molecules become weaker in solutions including 1-heptanol or 1-decanol. The main feature of 1-alkanol-aniline interactions is their sharper dependence with $n_{\text{OH}}$. For the sake of comparison, we also provide the $\Delta H_{\text{OH-NH2}}^{\text{int}}$ value for the methanol + piperidine mixture. It seems that interactions between unlike molecules in such solution are slightly weaker ($-36.9$ kJ·mol$^{-1}$) than those between methanol and cyclohexylamine ($-39.8$ kJ·mol$^{-1}$).

### 7.2. The $\Delta H_{\text{OH-OH}}$ term

This positive contribution depends on the considered solvent. For a given 1-alkanol, $H_{\text{m}}^{\text{E}}$(heptane) < $H_{\text{m}}^{\text{E}}$(cyclohexane) (cyclization effect) as it can be seen from the following experimental results. $H_{\text{m}}^{\text{E}}$(heptane)/J·mol$^{-1}$ = 591 ($n_{\text{OH}}$ = 2) [99]; 575 [100]; ($n_{\text{OH}}$ = 4); 575 [101]; ($n_{\text{OH}}$ = 5); 527 ($n_{\text{OH}}$ = 6) [102]; 427 ($n_{\text{OH}}$ = 10) [103] and $H_{\text{m}}^{\text{E}}$(cyclohexane)/J·mol$^{-1}$ = 624 ($n_{\text{OH}}$ = 2) [104]; 588 [105]; ($n_{\text{OH}}$ = 4); 598 [106] ($n_{\text{OH}}$ = 5); 604 [107] ($n_{\text{OH}}$ = 6); 666 [106] ($n_{\text{OH}}$ = 10). The differences $H_{\text{m}}^{\text{E}}$(cyclohexane)- $H_{\text{m}}^{\text{E}}$(heptane) become larger for systems formed by longer 1-alkanols. Similarly, we note that $H_{\text{m}}^{\text{E}}$(heptane) < $H_{\text{m}}^{\text{E}}$(toluene, isomeric molecule of aniline) (aromacity effect) since $H_{\text{m}}^{\text{E}}$(toluene)/J·mol$^{-1}$ [108] = 622 ($n_{\text{OH}}$ = 1); 881 ($n_{\text{OH}}$ = 3); 942 ($n_{\text{OH}}$ = 4); 912 ($n_{\text{OH}}$ = 5). These values clearly indicate that an aromatic hydrocarbon such as toluene is a more efficient breaker of the alcohol network.

### 7.3 The $\Delta H_{\text{NH2-NH2}}$ term

This contribution is positive and increases in line with $n_{\text{OH}}$, which can be ascribed to the larger aliphatic surfaces of longer 1-alkanols break more easily the amine-amine interactions. Note that $H_{\text{m}}^{\text{E}}$/J·mol$^{-1}$ of 1-hexylamine + $n$-alkane mixtures increases with the alkane size: 1064 (heptane); 1211 (decane); 1513 (hexadecane) [109]. The same occurs for the UCSTs of aniline systems: 342.6 (hexane) [110]; 343.1 (heptane) [111]; 356.8 (dodecane) [112] (all values in K). The existence of these miscibility gaps indicates that amine-amine interactions are much stronger in aniline systems. Accordingly, the $H_{m1}^{E,\infty}$ value of the aniline + heptane (15 kJ.mol$^{-1}$) [40,113] mixture is much higher than the results for mixtures 1-hexylamine + heptane ($H_{m1}^{E,\infty}$/kJ.mol$^{-1}$ = 5.7 [114]), or cyclohexylamine + cyclohexane ($H_{m1}^{E,\infty}$/kJ·mol$^{-1}$ = 5.5 [115]), characterized by rather similar $H_{m1}^{E,\infty}$ results.

### 7.4 The $\Delta H_{OH-NH2}$ term

This negative contribution can be roughly estimated from the product ($\Delta H_{OH-NH2}^{int}$ x number of interactions between unlike molecules created during mixing) (see, *e.g.*, [116]). Values of $\Delta H_{OH-NH2}^{int}$ have been already discussed (Table 8). One can expect that the second factor of the product decreases for larger $n_{OH}$ values, since the OH group is then more sterically hindered and a lower number of alcohol-amine interactions are formed upon mixing. In summary, the present contribution becomes less negative when $n_{OH}$ is increased.

### 7.5 Dependence of $H_m^E$ with the 1-alkanol and with the amine

The observed variation of $H_m^E$ with $n_{OH}$ for mixtures including aniline or cyclohexylamine can be explained taking into account that the three contributions to $H_m^E$ increase in line with $n_{OH}$, except for 1-hexylamine solutions, where the $\Delta H_{OH-OH}$ contribution decreases for longer 1-alkanols. In such a case, this effect is more or less counterbalanced with the increase of the $\Delta H_{NH2-NH2}$ and $\Delta H_{OH-NH2}$ terms and then $H_m^E$ slowly increases with $n_{OH}$.

For a given 1-alkanol, the much larger $H_m^E$ values of solutions with aniline can be ascribed to the $\Delta H_{OH-OH}$ and $\Delta H_{NH2-NH2}$ terms contribute largely to $H_m^E$. On the other hand, interactions are of dipolar type since, for these systems, $H_m^E$ is poorly described by the ERAS model [40].

Regarding mixtures with 1-hexylamine, or cyclohexylamine, both magnitudes, $\Delta H_{NH2-NH2}$ and $\Delta H_{OH-NH2}^{int}$, are practically independent of the amine, while the $\Delta H_{OH-OH}$ contribution is larger for systems containing cyclohexylamine and longer 1-alkanols. Thus, the more negative $H_m^E$ of systems including this amine and 1-alkanols with $n_{OH} \leq 4$ suggest that the $\Delta H_{OH-NH2}$ term must be more negative, probably due to the amine group is less sterically hindered and more interactions between unlike molecules are formed along the mixing process. In the case of systems with longer 1-alkanols, the difference $\left|\Delta H_{\text{OH-NH2}}(\text{cyclohexyalmine}) - \Delta H_{\text{OH-NH2}}(\text{1-hexylamine})\right|$ is lower than the difference $\left|\Delta H_{\text{OH-OH}}(\text{cyclohexyalmine}) - \Delta H_{\text{OH-OH}}(\text{1-hexylamine})\right|$, and the resulting $H_{\text{m}}^{\text{E}}$ values are very similar, and eventually slightly higher for the 1-decanol + cyclohexylamine mixture (Figure 4). Similar considerations are still valid to explain the larger $H_{\text{m}}^{\text{E}}$ result of the methanol + piperidine mixture ($-3160$ J·mol$^{-1}$ [117]).

For a binary mixture, the $S_{\text{CC}}(0)$ function is defined by [50, 51]:

$$S_{\text{CC}}(0) = \frac{RT}{(\partial^2 G_{\text{m}}^{\text{M}} / \partial x_1^2)_{P,T}} = \frac{x_1 x_2}{1 + \frac{x_1 x_2}{RT}\left(\frac{\partial^2 G_{\text{m}}^{\text{E}}}{\partial x_1^2}\right)_{P,T}} \tag{12}$$

For ideal mixtures, $G_{\text{m}}^{\text{E,id}} = 0$; and $S_{\text{CC}}^{\text{id}}(0) = x_1 x_2$. Stability conditions require that $S_{\text{CC}}(0) > 0$. Thus, if a system is close to phase separation, $S_{\text{CC}}(0)$ must be large and positive and the dominant trend is the separation between components (homocoordination), and $S_{\text{CC}}(0) > x_1 x_2$. If compound formation between components exists (heterocoordination), $S_{\text{CC}}(0)$ must be very low and $0 < S_{\text{CC}}(0) < x_1 x_2$. For more details, see reference [50]. Application of DISQUAC to calculate $S_{\text{CC}}(0)$ reveals that systems with 1-hexylamine or cyclohexylamine are characterized by heterocoordination, and that homocoordination is the dominant trend in aniline mixtures. For example, for 1-hexylamine mixtures, $S_{\text{CC}}(0)$ = 0.165 (methanol), 0.198 (1-pentanol); for cyclohexylamine solutions, $S_{\text{CC}}(0)$ = 0.120 (methanol), 0.128 (1-pentanol), and for aniline solutions, $S_{\text{CC}}(0)$ = 0.400 (methanol), 0.504 (1-pentanol). Moreover, these data show that, in mixtures with a given 1-alkanol, $S_{\text{CC}}(0)$ changes in the sequence: aniline > 1-hexylamine > cyclohexylamine. This confirms that a larger number of interaction between unlike molecules exists in solutions involving the cyclic amine.

*7.6    The interaction parameters*

Firstly, it is to be noted that the $\Delta H_{\text{OH-NH2}}^{\text{int}}$ (Table 8) and $\Delta h_{\text{AB}}^*$ (Table 6) values are quite similar, and it supports our calculations. On the other hand, the variation of the ERAS parameters with $n_{\text{OH}}$ for mixtures with cyclohexylamine is similar to those encountered when investigating other 1-alkanol + amine mixtures in terms of this model (Figures 6-8) [26,38,40]. The values of $\left|\Delta h_{\text{AB}}^*\right|$ and $\left|\Delta v_{\text{AB}}^*\right|$ are large since the present mixtures show large and negative $H_{\text{m}}^{\text{E}}$ and $V_{\text{m}}^{\text{E}}$ values (Figures 4 and 5) which arise from strong solvation effects. This means that the physical contributions to the excess functions, and therefore the physical parameter, are low as the following results for the methanol + cyclohexylamine system show. Thus, the chemical and physical contributions to $H_m^E$ are, respectively: ($-3815$ and $37$) J·mol$^{-1}$. For the $V_m^E$ function, the chemical and physical contributions are: ($-1.696$ and $-0.089$) cm$^3$·mol$^{-1}$. Similar results are obtained for the ethanol solution (Figure S3, supplementary material). The main difference between members of a homologous series is linked to their different $K_{AB}$ values (Figure 6). It is quite clear that the main solvation effects are encountered for the methanol system. We have determined the ERAS parameters for the methanol + piperidine mixture using data from reference [117] ($K_{AB} = 3500$; $\Delta h_{AB}^* = -39.6$ kJ·mol$^{-1}$; $\Delta v_{AB}^* = -9.8$ cm$^3$·mol$^{-1}$; $X_{AB} = 4$ J·cm$^{-3}$, and they fit well within the description provided above.

Regarding DISQUAC, we must underline some results. (i) The QUAC parameters are essentially the same for all the homologous series. A similar behaviour has been encountered for mixtures of the type 1-alkanol + *N,N*-dialkylamide [118], or + linear organic carbonate [119], or + cyclic ether [120]. (ii) In addition, they are very different to those of 1-alkanol + 1-hexylamine mixtures. Similarly, the $C_{hn,l}^{QUAC}$ (l =1,2,3) coefficients are different for methanol + di-*n*-propylamine or + piperidine systems [38,76] and this remarks that cyclic molecules are difficult to be treated using group contribution models [121,122].

## 8. Conclusions

Excess molar enthalpies have been measured for the systems methanol, ethanol, 1-propanol, 1-butanol, 1-heptanol, or 1-decanol + cyclohexylamine at 298.15 K and 0.1 MPa. The large and negative $H_m^E$ and $V_m^E$ values of these solutions reveal that they are essentially characterized by strong interactions between unlike molecules. Values of $U_{m,V}^E$ have been computed for 1-alkanol + cyclohexylamine, or + 1-hexylamine, or + aniline systems. The eos contribution to $H_m^E$ is particularly large for the methanol + aniline mixture, since $H_m^E$ and $U_{m,V}^E$ show opposite signs at equimolar composition. DISQUAC improves ERAS results on $H_m^E$. ERAS describes correctly the $V_m^E$ curves. The relative variation of $H_m^E$ of 1-alkanol + cyclohexyalmine, or + 1-hexylamine, or + aniline  mixtures with $n_{OH}$ in systems  with a given amine, or with the amine in mixtures with a given 1-alkanol has been discussed taking into account the different interactional contributions to $H_m^E$.

**CRediT authorship contribution statement**

L.F. Sanz, Conceptualization, Data Curation, Software, Validation, Original Draft. J.A. González, Conceptualization, Formal Analysis, Methology, Review and Editing. I. García de la Fuente, Investigation, Supervision, Original Draft, Writing. J.C. Cobos, Methodology, Investigation, Supervision, Validation. F. Hevia, Data Curation, Formal Analysis, Investigation, Writing

**TABLE 1**

Properties of pure compounds

| Compound | CAS | Purity[a] | $\rho$ [b]/g·cm$^{-3}$ | |
|---|---|---|---|---|
| | | | Exp. | Lit. |
| benzene | 71-43-2 | > 0.9995 | 0.873622 | 0.87360 [123] |
| cyclohexane | 110-82-7 | > 0.9999 | 0.773865 | 0.77366 [123] |
| 1-butanamine | 109-73-9 | > 0.9996 | 0.732758 | 0.73225 [124] |
| | | | | 0.73300 [125] |
| cyclohexanamine | 108-91-8 | > 0.999 | 0.862315 | 0.862207 [126] |
| methanol | 67-56-1 | > 0.9999 | 0.786716 | 0.78667 [127] |
| | | | | 0.7869 [128] |
| ethanol | 64-17-5 | > 0.9999 | 0.785086 | 0.7854 [128] |
| | | | | [0.78546 [129] |
| *n*-propan-1-ol | 71-23-8 | > 0.999 | 0.799770 | 0.79960 [130] |
| *n*-butan-1-ol | 71-36-3 | > 0.9986 | 0.805901 | 0.805762 [126] |
| | | | | 0.80575 [130] |
| *n*-heptan-1-ol | 111-70-6 | > 0.999 | 0.818987 | 0.81875 [131] |
| *n*-decan1-ol | 112-30-1 | > 0.987 | 0.826581 | 0.82644 [131] |

[a]value in mole fraction provided by the manufacturer (gas chromatograph analysis; [b]density at 298.15 K and 0.1 MPa. The standard uncertainties are: $u(T) = \pm 0.01$ K; $u(p) = \pm 1$ kPa; $u(\rho) = \pm 8 \times 10^{-5}$ g·cm$^{-3}$

**TABLE 2**

Experimental $H_m^E$ results at 298.15 K and 0.1 MPa for the systems used in the calibration of the microcalorimeter Tian-Calvet.[a]

| $x_1$ | $H_m^E$/J·mol$^{-1}$ | $x_1$ | $H_m^E$/J·mol$^{-1}$ |
|---|---|---|---|
| cyclohexane(1) + benzene (2)[b] | | methanol(1) + 1-butylamine(2)[c] | |
| 0.1054 | 349 | 0.0955 | −1051 |
| 0.2063 | 545 | 0.1680 | −1721 |
| 0.3056 | 686 | 0.2870 | −2718 |
| 0.4078 | 761 | 0.3977 | −3372. |
| 0.5095 | 802 | 0.4440 | −3583 |
| 0.5968 | 751 | 0.5181 | −3738 |
| 0.7061 | 689 | 0.5795 | −3771 |
| 0.7890 | 514 | 0.6597 | −3511 |
| 0.8895 | 311 | 0.6975 | −3338 |
| | | 0.7945 | −2548 |
| | | 0.9032 | −1306 |

[a]The standard uncertainties are: $u(T) = 0.01$ K, $u(p) = 1$ kPa, and $u(x_1) = 0.0005$. The relative combined expanded uncertainty (0.95 level of confidence) is $U_{rc}(H_m^E) = 0.03$; [b]coefficients from the fitting of the $H_m^E$ using equation (2): $A_0 = 3170$; $A_1 = 114$; $A_2 = 327$; $A_3 = -765$; $\sigma(H_m^E) = 14$ J.mol$^{-1}$ (equation 3); [c]coefficients from the fitting of the $H_m^E$ using equation (2): $A_0 = -14903$; $A_1 = -3492$; $A_2 = 2190$; $A_3 = 2696$; $\sigma(H_m^E) = 15$ J·mol$^{-1}$ (equation 3)

**TABLE 3.**

Excess molar enthalpies, $H_m^E$, at 298.15 K and 0.1 MPa for 1-alkanol(1) + cyclohexylamine(2) mixtures[a].

| $x_1$ | $H_m^E$/J·mol$^{-1}$ | $x_1$ | $H_m^E$/J·mol$^{-1}$ |
|---|---|---|---|
| methanol(1) + cyclohexylamine(2) | | ethanol(1) + cyclohexylamine(2) | |
| 0.1196 | −1263 | 0.0746 | −723 |
| 0.1574 | −1624 | 0.1514 | −1301 |
| 0.2181 | −2182 | 0.1614 | −1415 |
| 0.3253 | −3030 | 0.2364 | −1906 |
| 0.3906 | −3422 | 0.2886 | −2249 |
| 0.4494 | −3704 | 0.3961 | −2752 |
| 0.4993 | −3849 | 0.4913 | −3007 |
| 0.5491 | −3845 | 0.6021 | −2989 |
| 0.6003 | −3832 | 0.7040 | −2622 |
| 0.6974 | −3438 | 0.7561 | −2314 |
| 0.7965 | −2578 | 0.7964 | −2029 |
| 0.8452 | −2083 | 0.8520 | −1549 |
| 0.9005 | −1392 | 0.9122 | −990 |
| 1-propanol(1) + cyclohexylamine(2) | | 1-butanol(1) + cyclohexylamine(2) | |
| 0.1009 | −875 | 0.1003 | −835 |
| 0.1524 | −1296 | 0.1510 | −1206 |
| 0.1982 | −1612 | 0.2003 | −1553 |
| 0.2573 | −2016 | 0.2441 | −1842 |
| 0.3038 | −2290 | 0.3123 | −2226 |
| 0.3996 | −2726 | 0.4027 | −2598 |
| 0.5003 | −2948 | 0.5010 | −2810 |
| 0.6089 | −2855 | 0.6024 | −2755 |
| 0.6967 | −2573 | 0.6993 | −2397 |
| 0.7909 | −1975 | 0.7490 | −2153 |
| 0.8500 | −1467 | 0.8047 | −1751 |
| 0.8845 | −1183 | 0.8296 | −1580 |
| 0.9474 | −562 | 0.9062 | −918 |
| 1-heptanol(1) + cyclohexylamine(2) | | 1-decanol(1) + cyclohexylamine(2) | |
| 0.1012 | −738 | 0.1050 | −664 |
| 0.1496 | −1083 | 0.1535 | −959 |

TABLE 3 (continued)

| | | | |
|---|---|---|---|
| 0.1946 | −1373 | 0.2001 | −1228 |
| 0.2572 | −1740 | 0.2438 | −1467 |
| 0.3013 | −1974 | 0.3078 | −1792 |
| 0.3923 | −2338 | 0.4056 | −2132 |
| 0.5159 | −2581 | 0.5009 | −2303 |
| 0.6057 | −2510 | 0.6042 | −2258 |
| 0.7004 | −2213 | 0.7141 | −1942 |
| 0.7578 | −1919 | 0.7695 | −1676 |
| 0.7979 | −1666 | 0.8079 | −1458 |
| 0.8613 | −1216 | 0.8547 | −1135 |
| 0.8928 | −959 | 0.9058 | −784 |

[a]The standard uncertainties are: $u(T) = 0.01$ K, $u(p) = 1$ kPa, and $u(x_1) = 0.0005$. The relative combined expanded uncertainty (0.95 level of confidence) is $U_{rc}(H_m^E) = 0.03$.

**TABLE 4**

Coefficients $A_i$ and standard deviations, $\sigma(F_m^E)$ (equation (3)), for the representation of $F_m^E$ data at 298.15 K and 0.1 MPa for 1-alkanol(1) + cyclohexylamine(2) mixtures by equation (1).

| 1-alkanol | $F_m^E$ | $A_0$ | $A_1$ | $A_2$ | $A_3$ | $\sigma(F_m^E)/\text{J}\cdot\text{mol}^{-1}$ |
|---|---|---|---|---|---|---|
| methanol | $H_m^E$ | −15324 | −3649 | 2686 | 2164 | 18 |
|  | $U_{Vm}^E$ | −12748 | −3430 | 2315 | 1773 | 18 |
| ethanol | $H_m^E$ | −12033 | −2297 | 1426 | 1677 | 19 |
| 1-propanol | $H_m^E$ | −11768 | −1782 | 1959 | 1069 | 13 |
|  | $U_{Vm}^E$ | −9745 | −2303 | 1815 | 1093 | 13 |
| 1-butanol | $H_m^E$ | −11209 | −1639 | 1996 | 1005 | 11 |
|  | $U_{Vm}^E$ | −9364 | −1552 | 1697 | 916 | 11 |
| 1-heptanol | $H_m^E$ | −10260 | −1338 | 1904 |  | 13 |
|  | $U_{Vm}^E$ | −8859 | −1223 | 1624 |  | 13 |
| 1-decanol | $H_m^E$ | −9207 | −1414 | 1832 |  | 8 |
|  | $U_{Vm}^E$ | −8108 | −1152 | 1543 |  | 9 |

**TABLE 5**

Dispersive (DIS) and quasichemical (QUAC) interchange coefficients, $C_{hn,l}^{DIS}$ and $C_{hn,l}^{QUAC}$, for (h,n) contacts in 1-alkanol + cyclohexyalmine mixtures ($l = 1$, Gibbs energy; $l = 2$, enthalpy; $l = 3$, heat capacity)

| System | $C_{hn,1}^{DIS}$ | $C_{hn,2}^{DIS}$ | $C_{hn,3}^{DIS}$ | $C_{hn,1}^{QUAC}$ | $C_{hn,2}^{QUAC}$ | $C_{hn,3}^{QUAC}$ |
|---|---|---|---|---|---|---|
| methanol | 0.55 | −15.7 | 22[a] | −3.85 | −5 | 6[a] |
| ethanol | 2.5 | −15.7 | 22[a] | −3.85 | −3 | 6[a] |
| 1-propanol | 5 | −15.7 | 22[a] | −3.85 | −3 | 6[a] |
| 1-butanol | 5 | −16.7 | 22[a] | −3.85 | −3 | 6[a] |
| 1-pentanol | 5 | −19.5 | 22[a] | −3.85 | −3 | 6[a] |
| 1-heptanol | 5 | −19.5 | 22[a] | −3.85 | −3 | 6[a] |
| 1-decanol | 5 | −19.5 | 22[a] | −3.85 | −3 | 6[a] |

[a]guessed value

**TABLE 6**

ERAS parameters[a] for 1-alkanol(A) + cyclohexylamine(B) mixtures at 298.15 k and 0.1 MPa

| 1-alkanol | $K_{AB}$ | $\Delta h^*_{AB}$ /J·mol$^{-1}$ | $\Delta v^*_{AB}$ /cm$^3$·mol$^{-1}$ | $X_{AB}$ /J·mol$^{-3}$ |
|---|---|---|---|---|
| methanol | 3500 | −42.8 | −12.0 | 4 |
| ethanol | 2500 | −38.4 | −11.0 | 4 |
| 1-propanol | 1500 | −37.8 | −10.7 | 4 |
| 1-butanol | 1400 | −37.3 | −10.5 | 4 |
| 1-heptanol | 650 | −36.4 | −10.2 | 5 |
| 1-decanol | 300 | −36.4 | −10.0 | 7 |

[a] $K_{AB}$, equilibrium constant; $\Delta h^*_{AB}$, molar enthalpies of intermolecular hydrogen-bonding; $\Delta v^*_{AB}$, molar hydrogen-bonding volumes; $X_{AB}$, physical parameter

**TABLE 7**

Molar excess enthalpies, $H_m^E$, at equimolar composition, 298.15 K and 0.1 MPa of 1-alkanol + cyclohexylamine mixtures. Comparison with ERAS and DISQUAC results obtained using parameters from Tables 5 and 6.

| 1-alkanol | $H_m^E$ / J.mol$^{-1}$ | | | $\sigma(H_m^E)$ [a] /J.mol$^{-1}$ | | |
|---|---|---|---|---|---|---|
| | Exp. | ERAS | DQ | Exp. | ERAS | DQ |
| methanol | −3831 | −3778 | −3857 | 18 | 102 | 47 |
| ethanol | −3008 | −3034 | −3055 | 19 | 111 | 32 |
| 1-propanol | −2942 | −2934 | −2929 | 13 | 176 | 51 |
| 1-butanol | −2802 | −2779 | −2817 | 11 | 197 | 78 |
| 1-heptanol | −2565 | −2511 | −2577 | 13 | 275 | 140 |
| 1-decanol | −2302 | −2190 | −2308 | 8 | 325 | 160 |

[a]calculated using equation 2 with $F_{mcal,j}^E$ values determined from DISQUAC and ERAS models using interaction parameters from Tables 5 and 6, respectively.

**TABLE 8**

Partial molar excess enthalpies,[a] $H_{m1}^{E,\infty}$, at 298.15 K at 0.1 MPa for amine(1) + alkane(2) and for 1-alkanol(1) + amine(2) mixtures and hydrogen bond enthalpies, $\Delta H_{OH-NH2}^{int}$, for 1-alkanol- amine interactions.

| System | $H_{m1}^{E,\infty}$ /kJ·mol$^{-1}$ | $\Delta H_{OH-NH2}^{int}$ /kJ·mol$^{-1}$ |
|---|---|---|
| 1-hexylamine(1) + heptane(2) | 5.7 [114] | |
| cyclohexylamine(1) + cyclohexane(2) | 5.5 [115] | |
| piperidine(1) + cyclohexane(2) | 4.7 [132] | |
| aniline(1) + heptane(2) | 15.0 [40,113] | |
| methanol(1) + cyclohexylamine(2) | −11.1 | −39.8 |
| ethanol(1) + cyclohexylamine(2) | −10.0 | −38.7 |
| 1-propanol(1) + cyclohexylamine(2) | −9.1 | −37.8 |
| 1-butanol(1) + cyclohexylamine(2) | −8.6 | −37.3 |
| 1-heptanol(1) + cyclohexylamine(2) | −7.0 | −35.7 |
| 1-decanol(1) + cyclohexylamine(2) | −6.0 | −34.7 |
| methanol(1) + 1-hexylamine(2) | −9.7 [11] | −38.6 |
| 1-propanol (1) + 1-hexylamine(2) | −7.9 [11] | −36.8 |
| 1-pentanol(1) + 1-hexylamine(2) | −8.4 [11] | −37.3 |
| 1-octanol(1) + 1-hexylamine(2) | −8.1 [11] | −36.9 |
| 1-decanol(1) + 1-hexylamine(2) | −7.5 [11] | −36.4 |
| methanol(1) + aniline(2) | −0.04 [85] | −38.2 |
| ethanol (1) + aniline(2) | 2.5 [85] | −35.7 |
| 1-propanol(1) + aniline(2) | 3.9 [85] | −34.3 |
| 1-butanol(1) + aniline(2) | 5.6 [133] | −32.6 |
| 1-pentanol(1) + aniline(2) | 6.9 [85] | −31.3 |
| methanol(1) + piperidine(2) | −9.0 [117] | −36.9 |

[a]values obtained from $H_m^E$ data over the whole concentration range

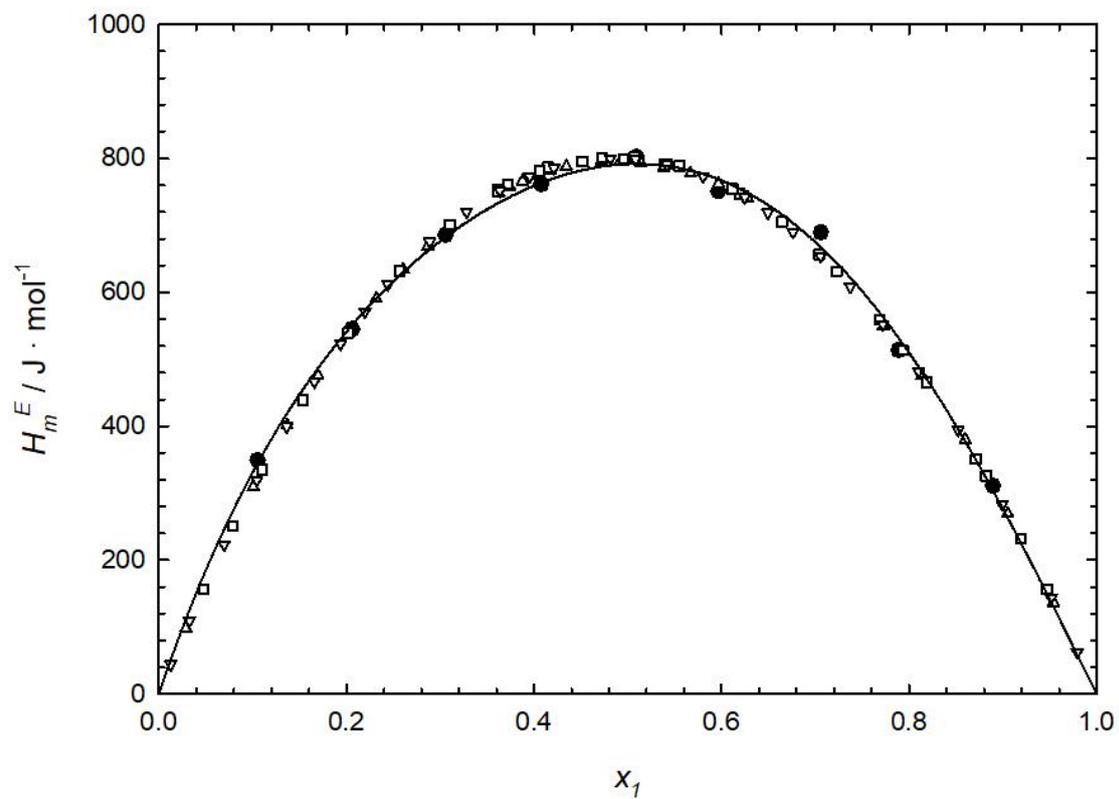

**Figure 1a.** $H_m^E$ of the cyclohexane(1) + benzene(2) system at 298.15 K and 0.1 MPa. Symbols, experimental results: (●), this work, (▽, [57]; (□), [58]; (△), [59]. Solid line: calculations with equation 2 using coefficients from Table 4.

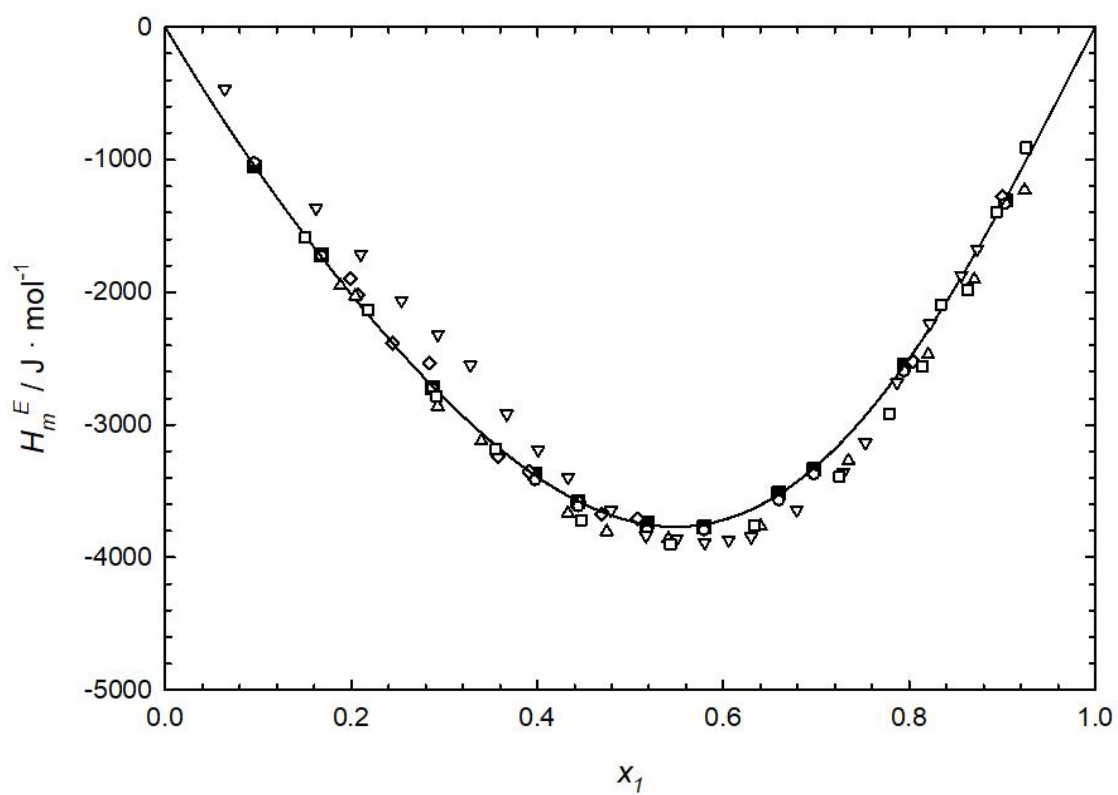

**Figure 1b.** $H_m^E$ of the methanol(1) + 1-butylamine(2) system at 298.15 K and 0.1 MPa. Symbols, experimental results: (■), this work, (◊); [16]; (○), [18]; (▽, [60]; (□), [61]; (△), [62]. Solid line: calculations with equation 2 using coefficients from Table 4.

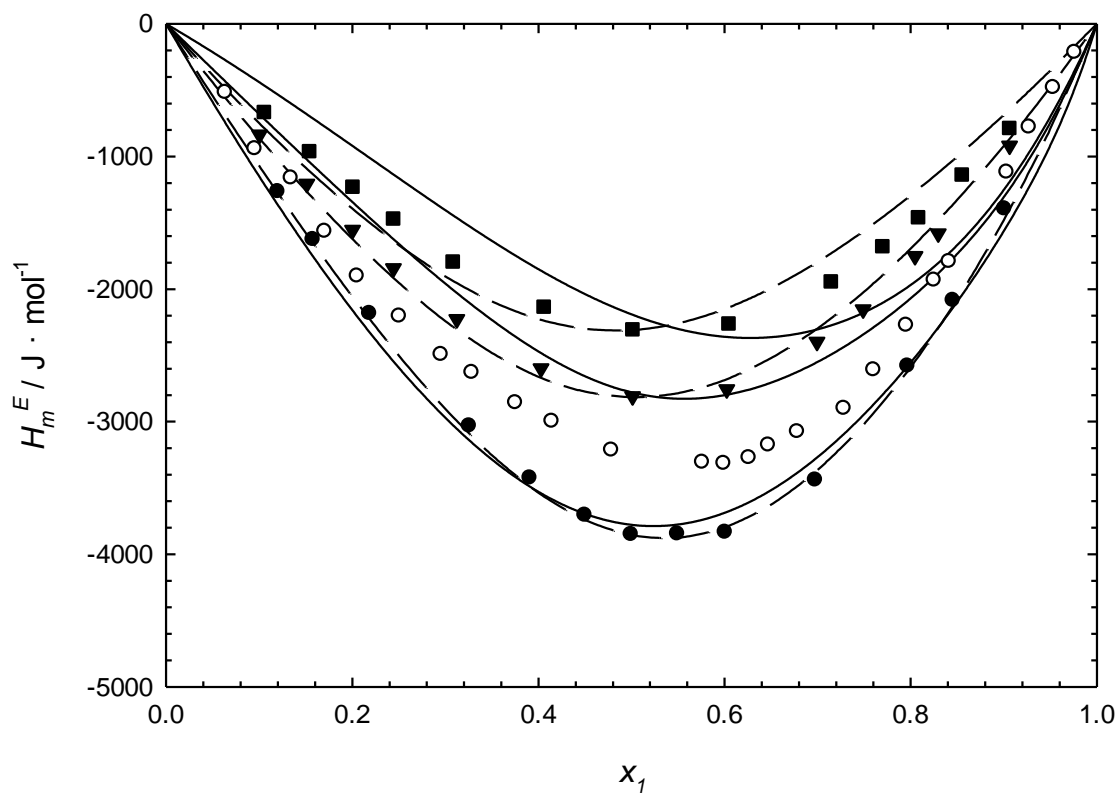

**Figure 2.** $H_m^E$ of 1-alkanol (1) + cyclohexylamine (2) systems at 298.15 K and 0.1 MPa. Symbols, experimental results: (●) methanol; (▼) 1-butanol; (■) 1-decanol (this work); (○) methanol [60]; Solid lines, ERAS results using parameters listed in Table 6. Dashed lines, DISQUAC calculations using interaction parameters from Table 5.

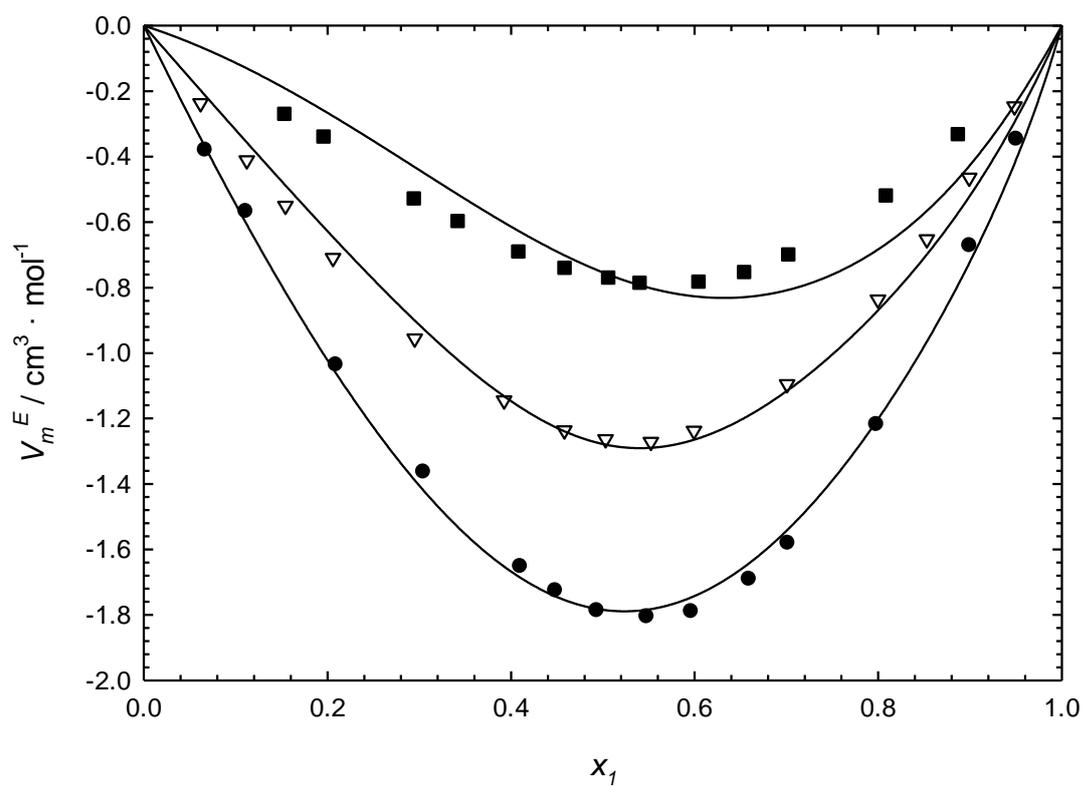

**Figure 3.** $V_m^E$ of 1-alkanol (1) + cyclohexylamine(2) systems at 298.15 K and 0.1 MPa. Symbols, experimental results: (●), methanol [29]; (▽), 1-butanol [27]; (■), 1-decanol [28]. Solid lines, ERAS results obtained with parameters from Table 6.

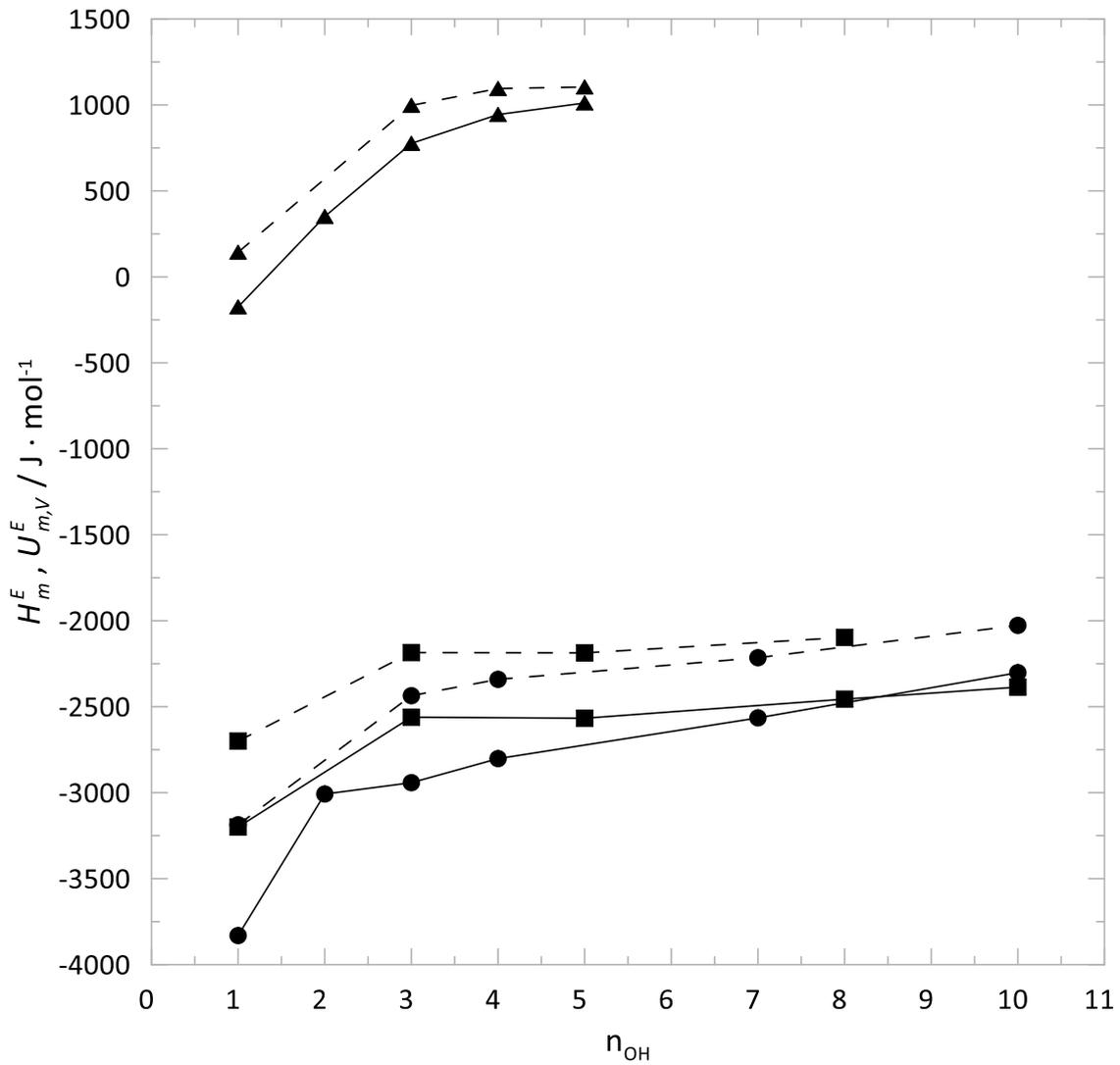

**Figure 4.** $H_m^E$ (solid lines) and $U_{m,V}^E$ (dashed lines, this work) at equimolar composition, $T = 298.15$ K and 0.1 MPa for 1-alkanol + amine mixtures vs. $n_{OH}$, the number of C atoms in the 1-alkanol: (■) 1-hexylamine [11]; (●) cyclohexylamine (this work); (▲) aniline [85,133].

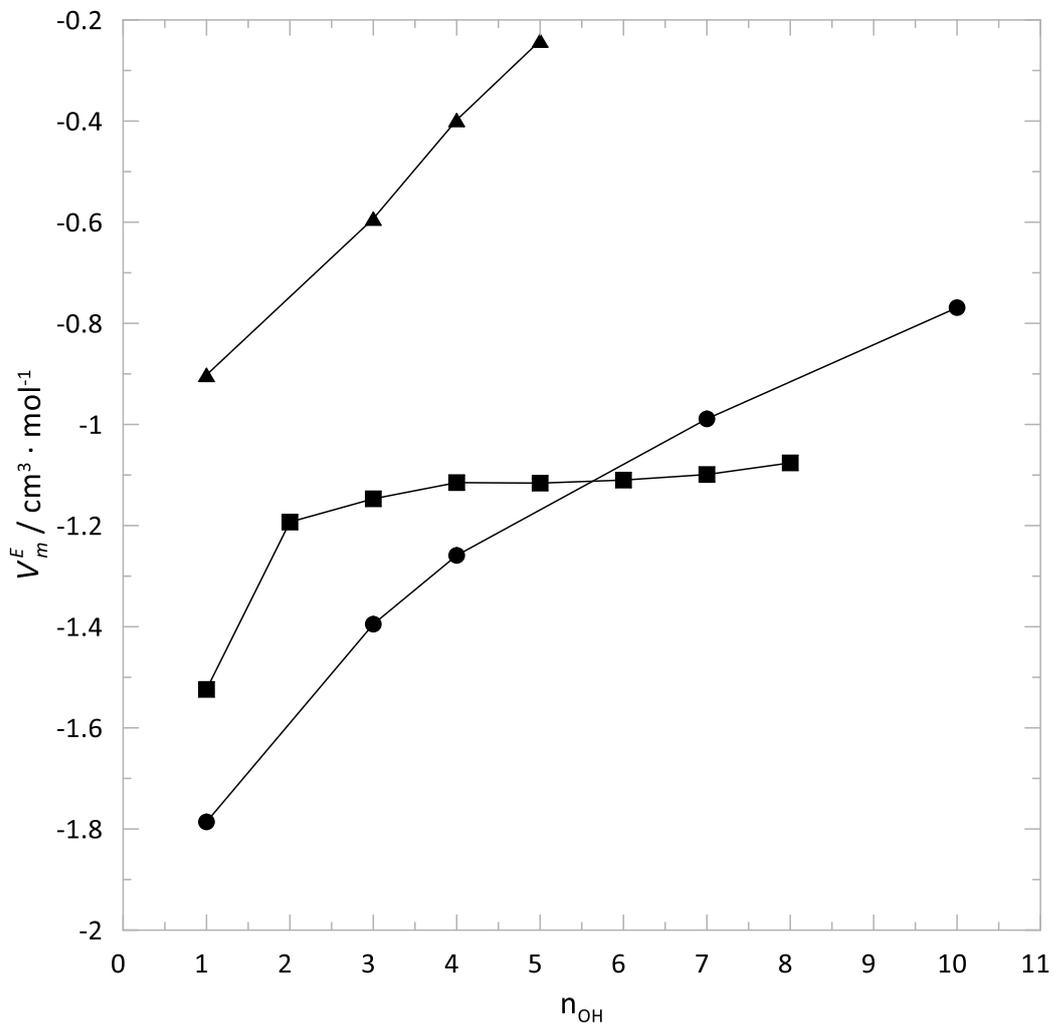

**Figure 5.** $V_m^E$ at equimolar composition, $T = 298.15$ K and 0.1 MPa for 1-alkanol + amine mixtures vs. $n_{OH}$, the number of C atoms in the 1-alkanol: (■) 1-hexylamine [26]; (●) cyclohexylamine [27-29]; (▲) aniline [30,134]

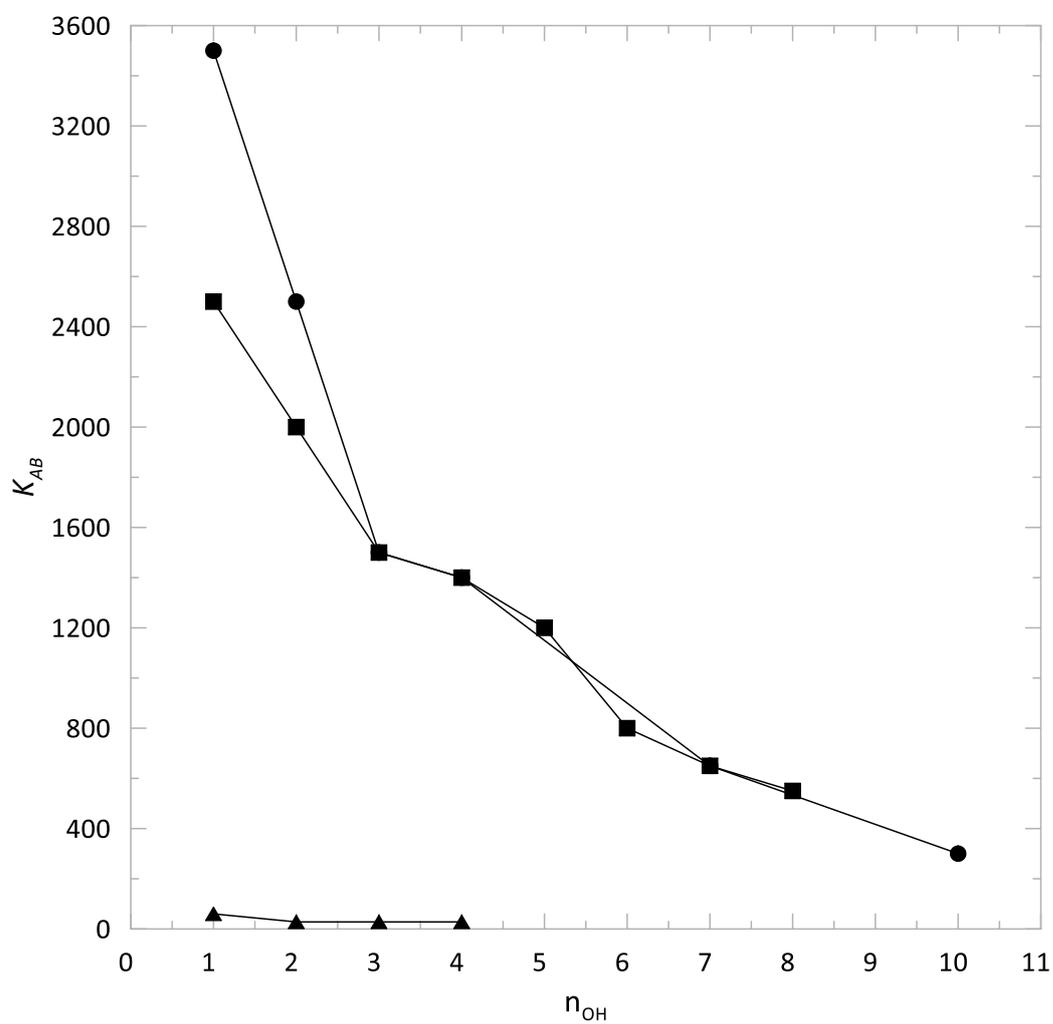

**Figure 6:** ERAS parameter $K_{AB}$ for 1-alkanol + amine systems at 298.15 K vs. $n_{OH}$, the number of C atoms in the 1-alkanol: (■), 1-hexylamine [26]; (●), cyclohexylamine (this work); (▲) aniline [40].

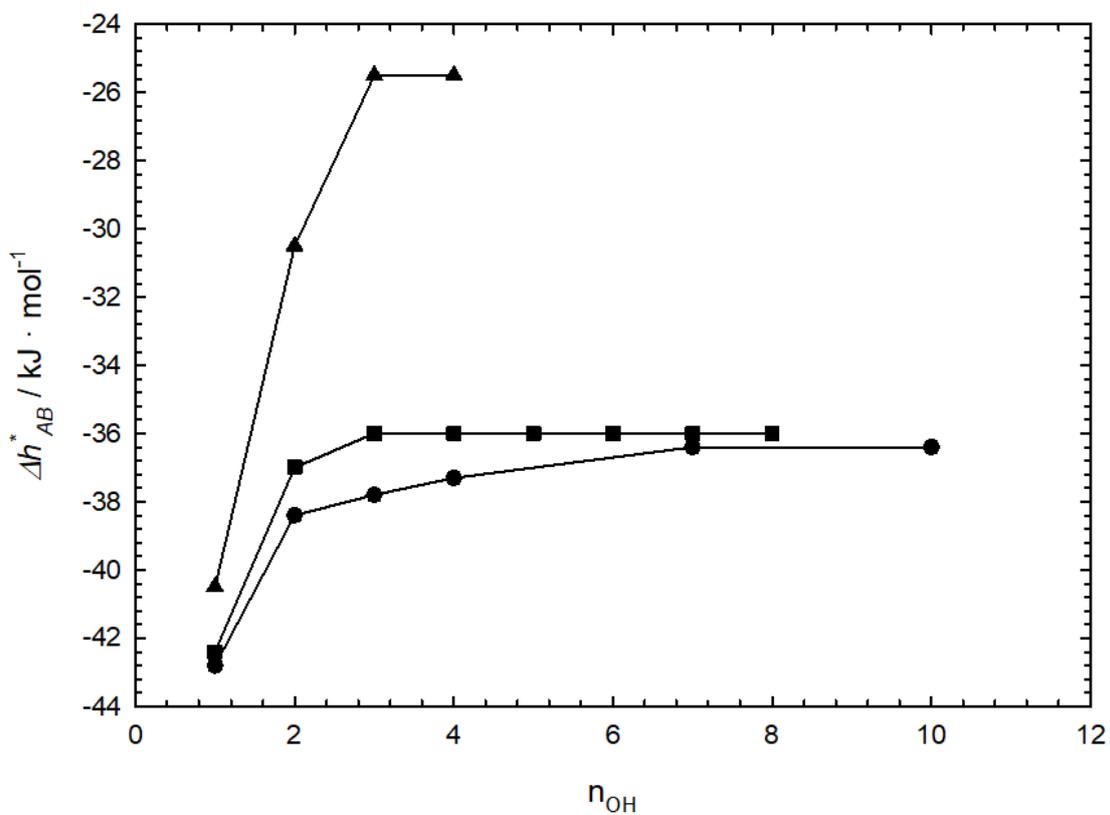

**Figure 7.** ERAS parameter $\Delta h^*_{AB}$ for 1-alkanol + amine systems vs. $n_{OH}$, the number of C atoms in the 1-alkanol: (■), 1-hexylamine [26]; (●), cyclohexylamine (this work); (▲) aniline [40].

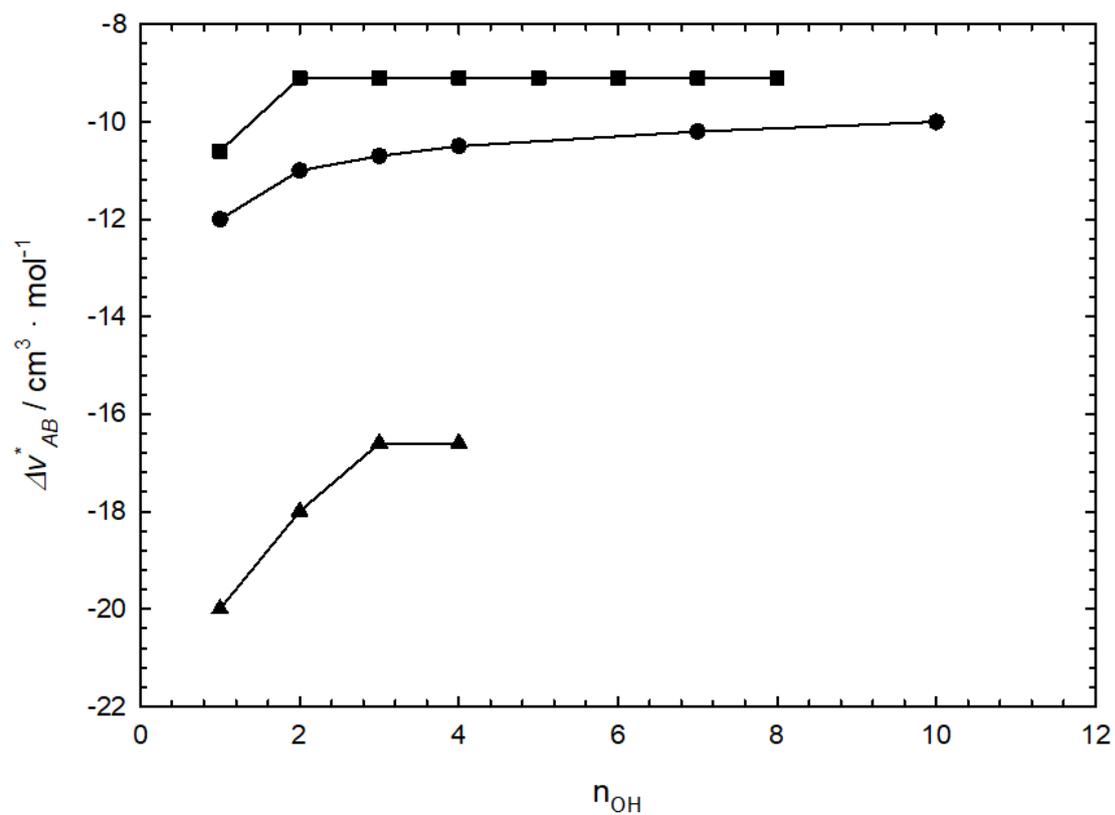

**Figure 8.** ERAS parameter $\Delta v^*_{AB}$ for 1-alkanol + amine systems vs. $n_{OH}$, the number of C atoms in the 1-alkanol: (■), 1-hexylamine [26]; (●), cyclohexylamine (this work); (▲) aniline [40].

SUPPLEMENTARY MATERIAL

**Thermodynamics of mixtures with strong negative deviations from Raoult's law. XVIII: Excess molar enthalpies for the (1-alkanol + cyclohexylamine) systems at 298.15 K and modelling**

Luis Felipe Sanz,[a] Juan Antonio González,[a*] Isaías. García de la Fuente,[a] José Carlos Cobos[a] and Fernando Hevia[b]

[a]G.E.T.E.F., Departamento de Física Aplicada, Facultad de Ciencias, Universidad de Valladolid, Paseo de Belén, 7, 47011 Valladolid, Spain.

[b]Université Clermont Auvergne, CNRS. Institut de Chimie de Clermont-Ferrand. F-63000, Clermont-Ferrand, France b Departamento de Física Aplacada. EIFAB. Campus D

*corresponding author, e-mail: jagl@termo.uva.es; Fax: +34-983-423136; Tel: +34-983-423757

**TABLE S1**

Excess molar internal energies at constant volume, $U_{m,V}^E$, at 298.15 K and 0.1 MPa for 1-alkanol(1) + cyclohexylamine(2) mixtures[a].

| $x_1$ | $U_{m,V}^E$ /J·mol$^{-1}$ | $x_1$ | $U_{m,V}^E$ /J·mol$^{-1}$ |
|---|---|---|---|
| Methanol(1) + cyclohexylamine(2) | | 1-propanol(1) + cyclohexylamine(2) | |
| 0.1196 | −1014 | 0.1009 | −663 |
| 0.1574 | −1308 | 0.1524 | −998 |
| 0.2181 | −1772 | 0.1982 | −1250 |
| 0.3253 | −2488 | 0.2573 | −1588 |
| 0.3906 | −2824 | 0.3038 | −1824 |
| 0.4494 | −3073 | 0.3996 | −2217 |
| 0.4993 | −3205 | 0.5003 | −2442 |
| 0.5491 | −3203 | 0.6089 | −2401 |
| 0.6003 | −3208 | 0.6967 | −2193 |
| 0.6974 | −2894 | 0.7909 | −1698 |
| 0.7965 | −2174 | 0.8500 | −1264 |
| 0.8452 | −1766 | 0.8845 | −1025 |
| 0.9005 | −1185 | 0.9474 | −490 |
| 1-butanol(1) + cyclohexylamine(2) | | 1-heptanol(1) + cyclohexylamine(2) | |
| 0.1003 | −687 | 0.1012 | −636 |
| 0.1510 | −992 | 0.1496 | −933 |
| 0.2003 | −1280 | 0.1946 | −1181 |
| 0.2441 | −1523 | 0.2572 | −1495 |
| 0.3123 | −1844 | 0.3013 | −1697 |
| 0.4027 | −2160 | 0.3923 | −2012 |

Table S1 (continued)

| | | | |
|---|---|---|---|
| 0.5010 | −2348 | 0.5159 | −2230 |
| 0.6024 | −2312 | 0.6057 | −2174 |
| 0.6993 | −2013 | 0.7004 | −1920 |
| 0.7490 | −1814 | 0.7578 | −1665 |
| 0.8047 | −1474 | 0.7979 | −1445 |
| 0.8296 | −1333 | 0.8613 | −1056 |
| 0.9062 | −776 | 0.8928 | −832 |
| 1-decanol(1) + cyclohexylamine(2) | | | |
| 0.1050 | −598 | | |
| 0.1535 | −858 | | |
| 0.2001 | −1094 | | |
| 0.2438 | −1303 | | |
| 0.3078 | −1588 | | |
| 0.4056 | −1881 | | |
| 0.5009 | −2028 | | |
| 0.6042 | −1986 | | |
| 0.7141 | −1706 | | |
| 0.7695 | −1472 | | |
| 0.8079 | −1279 | | |
| 0.8547 | −994 | | |
| 0.9058 | −688 | | |

[a]The standard uncertainties are: $u(T) = 0.01$ K, $u(p) = 1$ kPa, and $u(x_1) = 0.0005$. The relative combined expanded uncertainty (0.95 level of confidence) is $U_{rc}(U_{m,V}^{E}) = 0.06$.

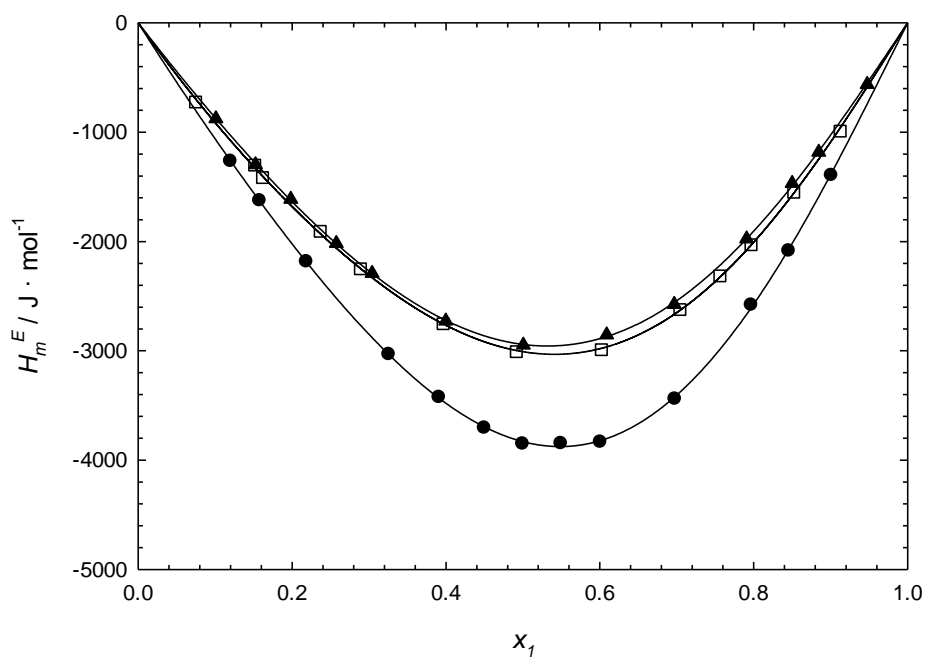

**Figure S1:** $H_\mathrm{m}^\mathrm{E}$ for 1-alkanol(1) + cyclohexylamine(2) systems at 298.15 K and 0.1 MPa. Symbols, experimental results: (●) methanol; (□) ethanol; (▲) 1-propanol. Solid lines: calculations from eq. (2) using coefficients from Table 4.

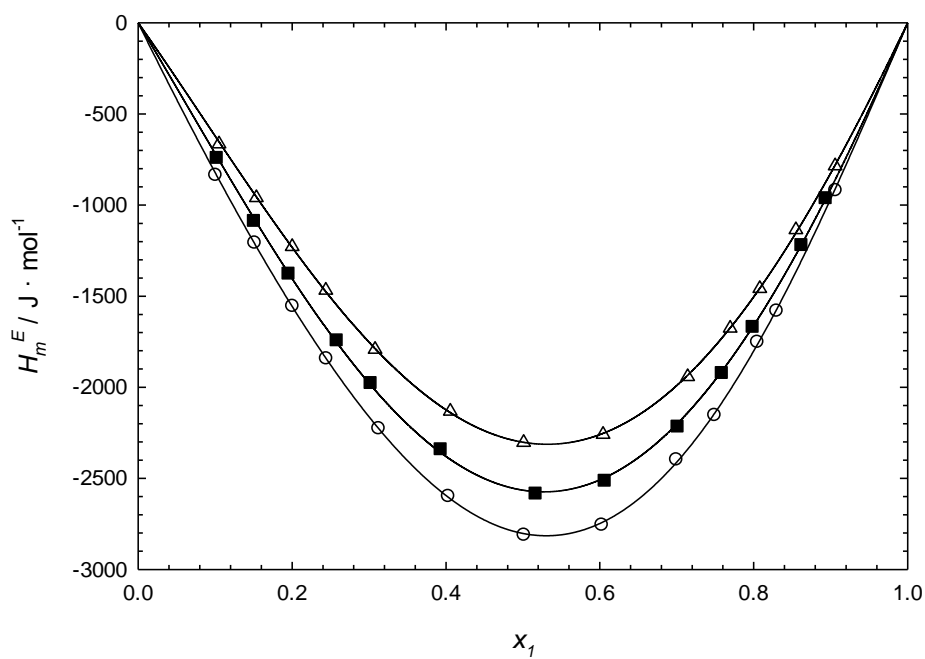

**Figure S2:** $H_m^E$ for 1-alkanol(1) + cyclohexylamine(2) systems at 298.15 K and 0.1 MPa. Symbols, experimental results: (○) 1-butanol; (■) 1-heptanol; (△) 1-decanol. Solid lines: calculations from eq. (2) using coefficients from Table 4.

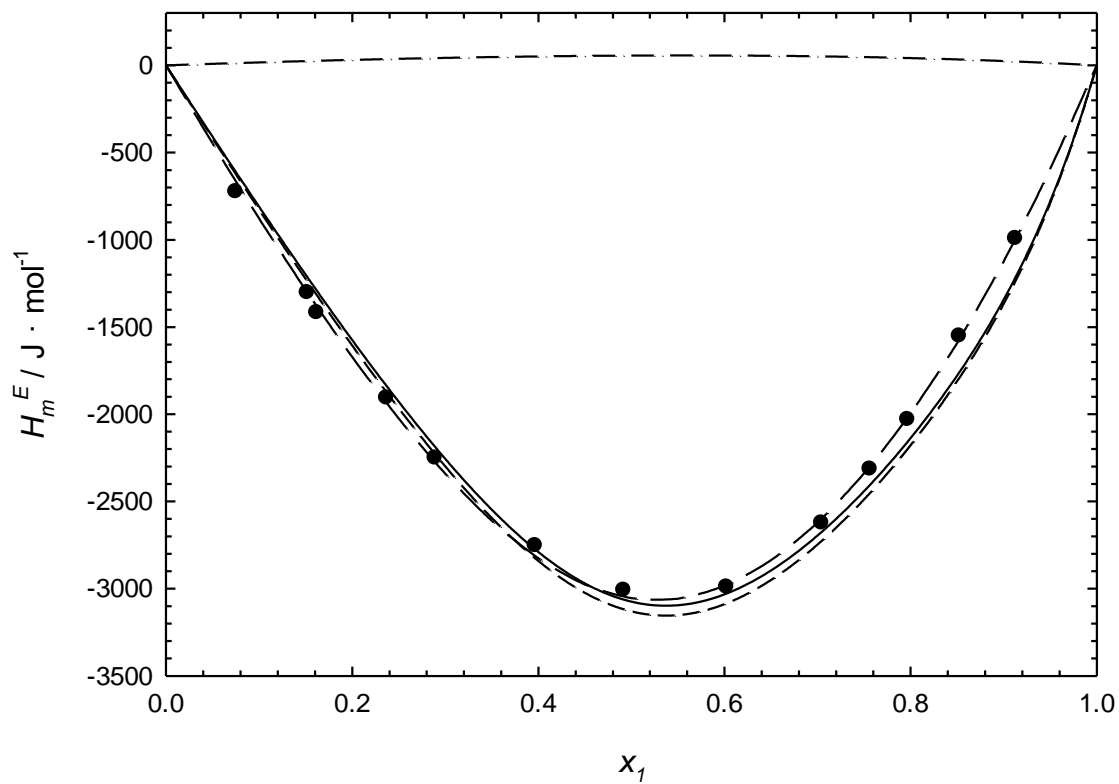

**Figure S3.** $H_m^E$ at 298.15 K and 0.1 MPa for the ethanol (1) + cyclohexylamine (2) system. Solid lines, ERAS results with parameters listed in Table 6; (-----), $H_{m,chem}^E$ contribution in ERAS; (-.-.-), $H_{m,phys}^E$ contribution in ERAS; (————), results from DISQUAC using interaction parameters listed in Table 5.

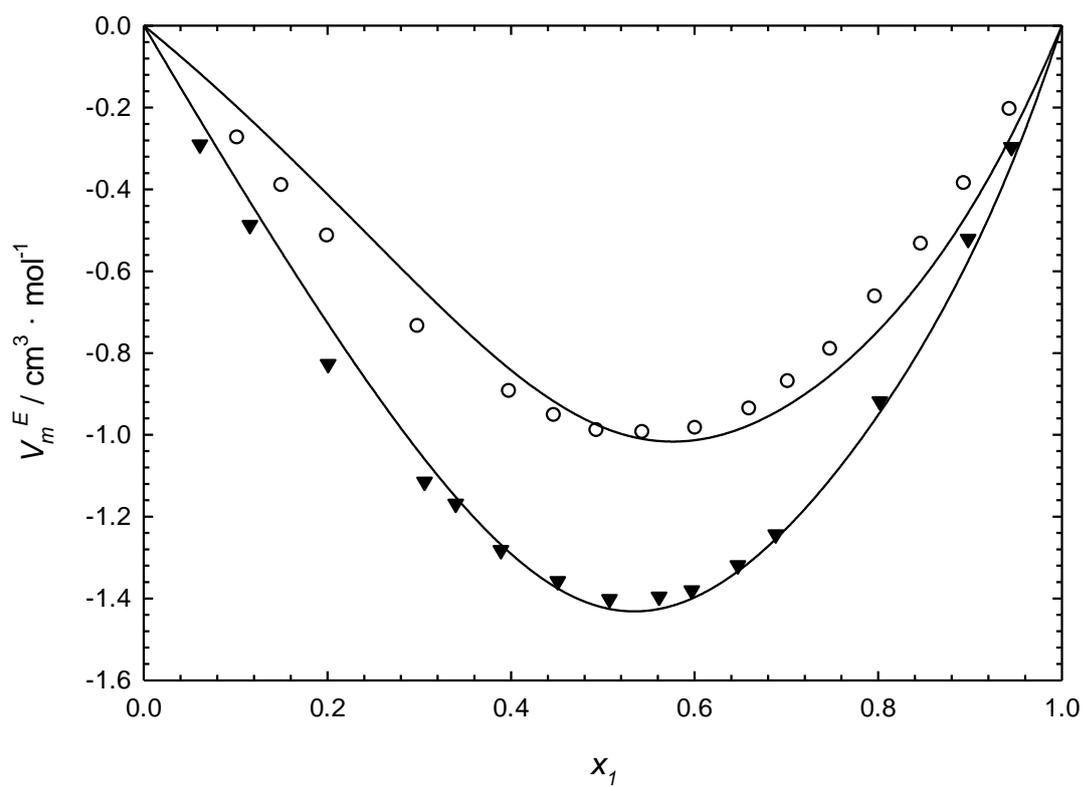

**Figure S4:** $V_\text{m}^\text{E}$ at 298.15 K and 0.1 MPa for 1-alkanol(1) + cyclohexylamine(2) systems. Symbols, experimental results: (▼) 1-propanol; (○) 1-heptanol. Solid lines, ERAS calculations with parameters from Table 6.